# An Ensemble of Convolutional Neural Networks for Audio Classification

Loris Nanni[a] Gianluca Maguolo[a] Sheryl Brahnam[b*], Michelangelo Paci[c]
[a] DEI, University of Padua, Viale Gradenigo 6, Padua, Italy. E-mail: {loris.nanni, gianluca.maguolo}@unipd.it
[b*] Department of Information Technology and Cybersecurity, Glass Hall, Room 387, Missouri State University, Springfield, MO 65804, USA. E-mail: sbrahnam@missouristate.edu.
[c] BioMediTech, Faculty of Medicine and Health Technology, Tampere University, Arvo Ylpön katu 34, D 219, FI-33520, Tampere, Finland. E-mail: michelangelo.paci@tuni.fi

**Abstract.** Research in sound classification and recognition is rapidly advancing in the field of pattern recognition. In this paper, ensembles of classifiers that exploit several data augmentation techniques and four signal representations for training Convolutional Neural Networks (CNNs) for audio classification are presented and tested on three freely available audio benchmark datasets: i) bird calls, ii) cat sounds, and iii) the Environmental Sound Classification (ESC-50). The best performing ensembles combining data augmentation techniques with different signal representations are compared and shown to either outperform or perform comparatively to the best methods reported in the literature on these datasets, including the challenging ESC-50 dataset. To the best of our knowledge, this is the most extensive study investigating ensembles of CNNs for audio classification. Results demonstrate not only that CNNs can be trained for audio classification but also that their fusion using different techniques works better than the stand-alone classifiers.

## 1. Introduction

Sound classification and recognition have long been included in the field of pattern recognition. Some of the more popular application domains include speech recognition [1], music classification [2], biometric identification [3], and environmental sound recognition [4]. Following the three classical pattern recognition steps of i) preprocessing, ii) feature/descriptor extraction, and iii) classification, most early work in sound classification began by extracting features such as the Statistical Spectrum Descriptor or Rhythm Histogram [5] from audio traces. Once it was recognized, however, that visual representations of audio, such as spectrograms [6] and Mel-frequency Cepstral Coefficients spectrograms [7], contain valuable information, powerful texture extraction techniques popular in image classification began to be investigated.

One of the first to investigate features from visual representations of audio was Costa et al., who in [8] computed gray level co-occurrence matrices (GLCMs) [9] from spectrograms as features to train Support Vector Machines (SVMs) on the Latin Music Database (LMD) [10] and in [11] Local Binary Patterns (LBPs) [12] to train SVMs on ISMIR04 [13]. Costa et al. [14] later investigated extracting Local Phase Quantization (LPQ) and Gabor filters [15]. Ensembles of classifiers designed to fuse a set of the most robust texture descriptors with acoustic features extracted from the audio traces on multiple datasets of texture descriptors were exhaustively investigated by Nanni et al. [2], who demonstrated that the accuracy of systems based solely on acoustic or visual features could be enhanced by combining many types of texture features.

Recently, deep learning classifiers have proven even more robust in pattern recognition and classification than have texture analysis techniques. Humphrey and Bello [16, 17] were among the first to apply CNNs to audio images for music classification. In the same year, Nakashika et al. [18] reported converting spectrograms to GCLM maps to train CNNs to performed music genre classification on the GTZAN dataset [19]. Later, Costa et al. [20] fused a CNN with the traditional pattern recognition framework of SVMs trained on LBP features to classify the LMD dataset. Additional advances requiring alterations in CNN structures honed specifically to address audio classification include the work of [21] and [22]. In [23], a multimodal system was produced that combined album cover images, reviews, and audio tracks for multi-label music genre classification.

The latest advances in deep learning have been applied to other sound recognition problems besides music genre recognition. For instance, biodiversity assessment via sound, which intends to monitor animal species at risk, has been enhanced by developments in animal and bird sound recognition. Some recent work relevant to biodiversity assessment includes [24] and [25]. In both works, the authors combined CNN with visual features to classify marine animals [26] and the sound of fish [27]. The fusion of CNNs with traditional techniques was shown to outperform both the traditional and single deep learning approaches.

Another important audio recognition problem has to do with identifying sources of noise in environments. This audio classification problem is of particular concern for cell phone developers since noise interferes with conversation. Consequently, datasets of extraneous sounds have been released to develop systems for handling different kinds of noise. The ESC-50 dataset, for instance, contains 2000 labeled samples divided into fifty classes of environmental sounds that range from dogs barking to the sound of sea waves and chainsaws. In [28], a deep CNN achieved results superior to human classification. Other work of interest in this area includes [29], [30], [31], [32], and [33]. For a more comprehensive survey of sound classification methods up to the present day, see [34].

For all its power, deep learning also has significant drawbacks when it comes to sound classification. For one, deep learning approaches require massive training data [35]. For audio classification, this means large numbers of labeled audio signals and their visual representations. Sound datasets are typically too small for deep learners. The process of developing sound datasets is prohibitively expensive and labor-intensive. There are methods for increasing the number of images in small datasets, however. One such method is to apply data augmentation techniques. Audio signals can be augmented in both time and frequency domains, and these augmentation techniques can be directly applied either on the raw signals themselves or on the images obtained after they have been converted into spectrograms. In [36], for example, several augmentation techniques were applied to the training set in the BirdCLEF 2018 dataset. The augmentation pipeline involved taking the original bird audio signals, chunking them, and then augmenting them in the time domain (e.g., by adding background/atmospheric noise) and frequency domain (e.g., by applying pitch shifts and frequency stretches). This augmentation process not only enlarged the dataset but also produced nearly a 10% improvement in identification performance. Similarly, some standard audio augmentation techniques such a time and pitch shifts for bird audio classification were applied in [37]. Samples were also generated in [37] by summing separate samples belonging to the same class. This summing technique was used for domestic sound classification in [38] and [39]. In [40], new data was generated by computing the weighted sum of two samples belonging to different classes and by teaching the network to predict the weights of the sum. Audio signal augmentation on a domestic cat sound dataset was produced in [41] by randomly time stretching, pitch shifting, compressing the dynamic range, and inserting noise. Data augmentation techniques that are standard in speech recognition have also proven beneficial for animal sound identification, as in [42] and [43].

The goal of this work is to investigate multiple sets of different data augmentation approaches and methods for representing an audio signal as an image, with each augmentation method combined with a separate CNN. Building such ensembles is motivated by two observations: 1) it is well known that ensembles of neural networks generally perform better than stand-alone models due to the instability of the training process [44], and 2) it has been shown in other classification tasks that an ensemble of multiple networks trained with different augmentation protocols performs much better than do stand-alone networks [45]. The scores of the neural networks trained here are combined by sum rule, and the proposed approach is tested across three different audio classification datasets: domestic cat sound classification ([41]), bird call classification [46], and environmental classification [4]. Our experiments were designed to compare methods with the aim of maximizing performance by varying sets of data augmentation methods with different image representations of the audio signals.

The main contribution of this study is the exhaustive tests performed on ensembles fusing CNNs trained with different data augmentation and signal representation combinations with their performance compared across the three datasets. Another contribution of this work is the free availability of the MATLAB code used in this study, available at https://github.com/LorisNanni.

## 2. Audio Image Representation

Since the input to a CNN is in the form of a matrix, the following four methods were used to map the audio signals into spectrograms:

1. The Discrete Gabor Transform (DGT): this is a Short-Time Fourier Transform (STFT) with a Gaussian kernel as the window function. The continuous version of DGT can be defined as the convolution between the product of the signal with a complex exponential and a Gaussian, as

$$G(\tau, \omega) = \frac{1}{\sigma^2} \int_{-\infty}^{+\infty} x(t) e^{i\omega t} e^{-\pi \sigma^2 (t-\tau)^2} \, dt, \qquad (1)$$

where $x(t)$ is the signal, $\omega$ is a frequency, and $i$ is the imaginary unit. The width of the Gaussian window is defined by $\sigma^2$. The discrete version of DGT applies the discrete convolution rather than the continuous convolution. The output $G(\tau, \omega)$ is a matrix, where the columns represent the frequencies of the signal at a fixed time. The DGT implementation used in this study (see, [47]) is available at http://ltfat.github.io/doc/gabor/sgram.html.

2. Mel spectrograms (MEL) [48]: these spectrograms are computed by extracting the coefficients relative to the compositional frequencies with STFT. Extraction is accomplished by passing each frame of the frequency-domain representation through a Mel filter bank (the idea is to mimic the non-linear human ear perception of sound, which discriminates lower frequencies better than higher frequencies). Conversion between Hertz (f) and Mel (m) is defined as

$$m = 2595 \log_{10}(1 + 700f). \qquad (2)$$

The filters in the filter bank are all triangular, which means that each has a response of 1 at the center frequency, which decreases linearly towards 0 until it reaches the center frequencies of the two adjacent filters, where the response is 0.

3. Gammatone (GA) band-pass filters: this is a bank of GA filters whose bandwidth increases with the increasing central frequency. The functional form of Gammatone is inspired by the response of the cochlea membrane in the inner ear of the human auditory system [49]. The impulse response of a Gammatone filter is the product of a statistical distribution (Gamma) and a sinusoidal carrier tone. This response can be defined as

$$h_i(t) = \begin{cases} a \cdot t^{n-1} e^{-2\pi B_i t} \cos(2\pi \omega_i t + \phi), & t \geq 0 \\ 0, & t < 0 \end{cases} \qquad (3)$$

where $\omega_i$ is the central frequency of the filter and $\phi$ its phase. Gain is controlled by the constant $a$, and $n$ is the order of the filter. $B_i$ is a decay parameter that determines the bandwidth of the band-pass filter.

4. Cochleagram (CO): this mapping models the frequency selectivity property of the human cochlea [50]. To extract a cochleagram, it is first necessary to filter the original signal with a gammatone filter bank (see 3 above). The filtered signal must then be divided into overlapping windows. For each window and every frequency, the energy of the signal is calculated.

Each of the four spectrograms is then mapped to a gray-scale image using a linear transformation that maps the minimum value to 0 and the maximum value to 255, with the value of each pixel rounded to the closest smaller integer.

## 3. Convolutional Neural Networks

CNNs are used for two different purposes in this study: 1) as a feature extractor, where the features are used to train simpler SVMs, and 2) as a classifier. Aside from the input and output layers, CNNs are composed of one or more of the following hidden layers: convolutional (CONV), activation (ACT), pooling (POOL), and fully-connected (FC), or classification layer. The CONV layers extract features from the input volume and work by convolving a local region of the input volume (the receptive field) to filters of the same size. Once the convolution is computed, these filters slide into the next receptive field, where once again, the convolution between the new receptive field and the same filter is computed. This process is iterated over the entire input image, whereupon it produces the input for the next layer, a non-linear ACT layer, which improves the learning capabilities and classification performance of the network. Typical activation functions include i) the nonsaturating ReLU function $f(x) = \max(0, x)$, ii) the saturating hyperbolic tangent $f(x) = \tanh(x)$, $f(x) = |\tanh(x)|$, and iii) the sigmoid function $f(x) = (1 + e^{-x})^{-1}$. Pool layers are often interspersed between CONV layers and perform non-linear downsampling operations (max or average pool) that serve to reduce the spatial size of the representation, which in turn has the benefit of reducing the number of parameters, the possibility of overfitting, and the computational complexity of the CNN. FC layers typically make up the last hidden layers and have fully connected neurons to all the activations in the previous layer. SoftMax is generally used as the activation function for the output CLASS layer, which performs the final classification (also typically using the SoftMax function).

In this study, five CNNs pretrained on ImageNet [51] or Places365 [52] are adapted to the problem of sound classification as defined in the datasets used in this work. The architecture of the following pre-trained CNNs remains unaltered except for the last three layers, which are replaced by an FC layer, an ACT layer using SoftMax, and a CLASS layer also using SoftMax:

1. AlexNet [53] is the first neural network to win (and by a large margin) the ILSVRC 2012 competition. AlexNet has a structure composed of five CONV blocks followed by three FC layers. The dimension of the hidden layers in the network is gradually reduced with max-pooling layers. The architecture of AlexNet is simple since every hidden layer has only one input layer and one output layer.

2. GoogleNet [54] is the winner of ILSVRC 2014 challenge. The architecture of GoogleNet involves twenty-two layers and five POOL layers. GoogleNet was unique in its introduction of a novel *Inception* module, which is a subnetwork made up of parallel convolutional filters. Because the output of these filters is concatenated, the number of learnable parameters is significantly reduced. This study uses two pre-trained GoogleNets: the first is trained on the ImageNet database [51], and the second is trained on the Places365 [52] datasets.

3. VGGNet [55] is a CNN that took second place in ILSVRC 2014. Because VGGNet includes 16 (VGG-16) or 19 (VGG-19) CONV/FC layers, it is considered extremely deep. All the CONV layers are homogeneous. Unlike AlexNet [53], which applies a POOL layer after every CONV layer, VGGNet is composed of relatively tiny $3 \times 3$ convolutional filters with a POOL layer applied every two to three CONV layers. Both VGG-16 and VGG-19 are used in this study, and both are pre-trained on the ImageNet database [51].

4. ResNet [56] is the winner of ILSVRC 2015 and is much deeper than VGGNet. ResNet is distinguished by introducing a novel *network-in-network* architecture composed of residual (RES) layers. ResNet is also unique in applying global average pooling layers at the end of the network rather than the more typical set of FC layers. These architectural advances produce a model that is eight times deeper than VGGNet yet significantly smaller in size. Both ResNet50 (a 50 layer Residual Network) and ResNet101 (the deeper variant of ResNet50) are investigated in this study. Both CNNs have an input size 224×224 pixels.

5. InceptionV3 [57] advances GoogleNet by making the auxiliary classifiers perform as regulators rather than as classifiers. This is accomplished by factorizing 7x7 convolutions into two or three consecutive layers of 3×3 convolutions and applying the RMSProp Optimizer. InceptionV3 accepts images of size 299×299 pixels.

## 4. Data Augmentation approaches

Below is a description of the augmentation protocols that were combined and tested in this study. For each data augmentation method used to train a CNN, both the original and the artificially generated patterns were included in the training set.

### 4.1 Standard Signal Augmentation (SGN)

SGN is the application of the built-in data augmentation methods for audio signals available in MATLAB. For each training signal, ten new ones were generated by applying the following labeled transformations with 50% probability:

1. *SpeedupFactorRange* scales the speed of the signal by a random number in the range of [0.8, 1.2];

2. *SemitoneShiftRange* shifts the pitch of the signal by a random number in the range of $[-2,2]$ semitones;

3. *VolumeGainRange* increases or decreases the gain of the signal by a random number in the range of $[-3,3]$ dB;

4. *SNR* injects random noise into the signal in the range of $[0, 10]$ dB;

5. *TimeShiftRange* shifts the time of the signal in the range of $[-0.005, 0.005]$s.

### 4.2 Short Signal Augmentation (SSA)

SSA works directly on the raw audio signals. For every original image, the following ten augmentations are applied to produce ten new images:

1. *applyWowResampling* implements wow resampling, a variant of pitch shift that changes the intensity in time. The formula for the wow transformation is

$$F(x) = x + a_m \frac{\sin(2\pi f_m x)}{2\pi f_m},$$

where x is the input signal. In this study, $a_m = 3$ and $f_m = 2$;

2. *applyNoise* is the insertion of white noise so that the ratio between the signal and the noise is $X$ dB; in this study $X = 10$;

3. *applyClipping* normalizes the audio signal by leaving 10% of the samples out of [-1, 1], with the out-of-range samples (**x**) clipped to sign(**x**).

4. *applySpeedUp* not only increases but also decreases the speed of the audio signal; in this study, the speed was augmented by 15%.

5. *applyHarmonicDistortion* is the repeated application of quadratic distortion to the signal; in this study, the following distortion was applied five consecutive times:

$$s_{out} = \sin(2\pi s_{in});$$

6. *applyGain* increases the gain by a specific number of dB, which in this study was set to ten dB;

7. *applyRandTimeShift* randomly divides each audio signal in two and swaps them by mounting them back into a randomly shifted signal. If we call $s_{in}(t)$ the value of the input audio signal at time $t$, $T$ is the length of the signal and $t^*$ is a random time between 0 and $T$:

$$s_{out}(t) = s_{in}(mod(t^* + t, T));$$

8. *applyDynamicRangeCompressor* applies Dynamic Range Compression (DRC) [58] to a sample audio signal. DRC boosts the lower intensities of an audio signal and attenuates the higher intensities by applying an increasing piecewise linear function. DRC, in other words, compresses an audio signal's dynamic range;
9. *applyPitchShift* shifts the pitch of an audio signal by a specific number of semitones. We chose to increase it by two semitones;
10. We use *applyPitchShift* again to decrease the pitch of the audio signal by two semitones.

**4.3 Super Signal Augmentation (SSiA)**

With this protocol, twenty-nine new images are generated from every original image. The following five augmentations are applied to every sample, with the parameters of the augmentations randomized to generate the new images:

1. *applyWowResampling,* as in SSA;
2. *applySpeedUp*, as in SSA; but, in this case, the speed is either increased or decreased by a random number of percentage points in the range $[-5, 5]$;
3. *applyGain*, as in SSA, but the gain factor is sampled randomly in the range of $[-0.5, 0.5]$;
4. *applyRandTimeShift*, as in SSA;
5. *applyPitchShift*, as in SSA, but the pitch is shifted in the range of $[-0.5, 0.5]$.

Small parameters are selected because applying multiple transformations to the input introduces changes that are too large. The difference between the protocols in SSiA and in SSA is that SSiA protocols create a large number of images through multiple small transformations. Conversely, the images created by SSA protocols are generated with only one large transformation.

### 4.4 Time Scale Modification (TSM)

This protocol applies the five algorithms contained in the TSM toolbox [59]. TSM methods are commonly used in music production software to change the speed of signals without changing their pitch. Since two different constant stretching factors (0.5 and 1.8) were used for each TSM method, this augmentation approach produced ten new images. For a detailed description of the TMS toolbox, see [60]. A brief description of the five TMS algorithms follows (see Figure 1 for some examples):

1. OverLap Add (OLA): this algorithm is the simplest TSM method. It covers the input signals with overlapping windows of size $H_a$ and maps them into overlapping windows of size $H_s$. The number $H_\alpha$ depends on the implementation of the algorithm, while the ratio $\alpha = H_s/H_a$ is the speed-up factor, which the user can optionally set. The settings investigated this study were 0.8 and 1.5. These same values were used for each TMS method.

2. Waveform Similarity OverLap Add (WSOLA): this is a modification of OLA where the overlap of the windows is not fixed but has some tolerance to better represent the output signal in cases where there is a difference of phase.

3. Phase Vocoder addresses the same phase problem as WSOLA. However, it exploits the dual approach by matching the windows in the frequency domain: first, the Fourier transforms of the signal are calculated; second, the frequencies are matched, and the signal is pulled back into the time domain.

4. Phase Vocoder with identity phase locking: this TSM method is a modification of Phase Vocoder where the frequencies are matched as if they were not independent of each other. This modification was introduced by Laroche and Dolson [61].

5. Harmonic-Percussive Source Separation (HPSS): this augmentation technique decomposes an audio signal into its harmonic sound components, which form structures in the time direction, and its percussive sounds, which yield structures in the frequency direction. After decomposing the signal in this way, the phase vocoder is applied with the identity phase locking to the harmonic component, and OLA is applied to the percussive component. Finally, these two components are merged to form a new signal.

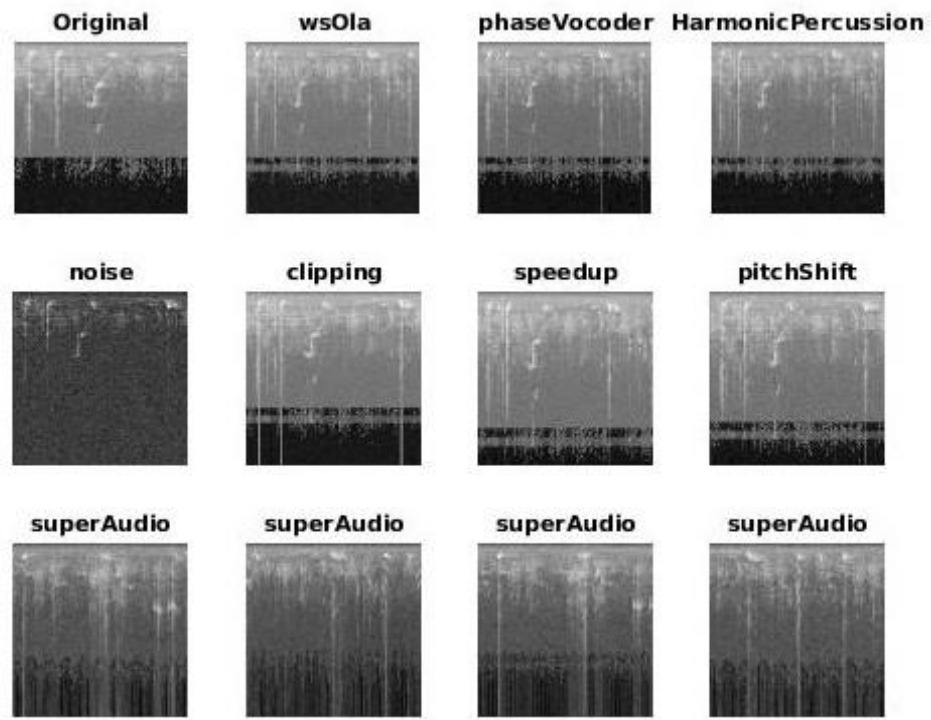

**Figure 1**. Audio augmented samples, where the last row shows the iterative applications of multiple random augmentations (SSiA)

### 4.5 Short Spectrogram Augmentation (SSpA)

SSpA works directly on spectrograms and generates five transformed versions of each original:

1. *applySpectrogramRandomShifts* randomly applies pitch shift and time shift.

2. *Vocal Tract Length Normalization* (apply*VTLN*) creates a new image by applying VTLP [42], which divides a given spectrogram into ten unique temporal slices. Once so divided, each slice passes through the following transformation:

$$G(f) = \begin{cases} \alpha f, & 0 \leq f < f_0 \\ \dfrac{f_{max} - \alpha f_0}{f_{max} - f_0}(f - f_0) + \alpha f_0, & f_0 \leq f \leq f_{max} \end{cases},$$

where $f_0, f_{max}$ are the basic and maximum frequency, and $\alpha \in [a, b]$ is randomly chosen. In this study, $a$ and $b$ are set to 0.9 and 1.1, respectively.

3. *applyRandTimeShift* does as its name indicates by randomly picking the shift value $T$ in $[1, M]$, where $M$ is the horizontal size of the input spectrogram. A given spectrogram is cut into two different images: $S_1$ and $S_2$, the first taken before and second after time $T$. The new image is generated by inverting the order of $S_1$ and $S_2$.

4. *applyRandomImageWarp* creates a new image by applying Thin-Spline Image Warping (TPS-Warp) [62] to a given spectrogram. TPS-Warp is a perturbation method applied to the original image by randomly changing the position of a subset $S$ of the input pixels. It adapts pixels that do not belong to $S$ via linear interpolation. In this study, the spectrogram is changed on the horizontal axis only. Also, a frequency-time mask is applied by setting to zero the values of two rows and one column of the spectrogram. In this study, the width of the rows is set to 5 pixels and the width of the column to 15 pixels.

5. *applyNoiseS* applies pointwise random noise to spectrograms. The value of a pixel is multiplied by a uniform random variable of average one and variance one, with probability 0.3.

**4.6 Super Spectro Augmentation (SuSA)**

In this protocol, twenty-nine new images are generated from each original sample. The following five augmentation methods are applied to each signal, with parameters randomized to produce different samples (see Figure 2 for some examples):

1. *applySpectrogramRandomShifts* as in SSpA, but with the time shift equal to zero and random pitch shift in the range $[-1, 1]$,

2. *applyVTLN* as in SSpA,

3. *applyRandTimeShift* as in SSpA,

4. *applyFrequencyMasking* sets to zero at most two random columns (which represent times) and at most two random rows (which represent frequencies),

5. *applyNoiseS* applies pointwise random noise to spectrograms. The value of a pixel is multiplied by a uniform random variable in $[0.3, 1.7]$, with a probability $0.1$.

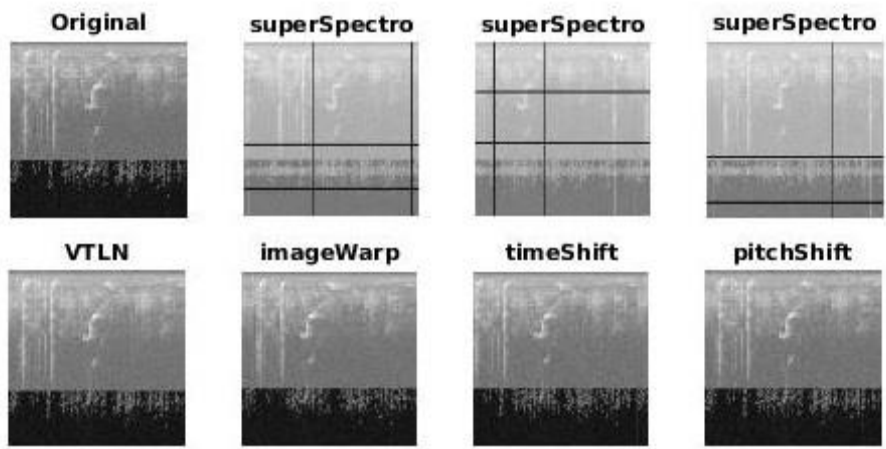

**Figure 2**. Spectrogram augmented samples.

## 5. Experimental results

The approach presented here was tested on three sound datasets:

- BIRDZ [46]: a control audio dataset, where the real-world recordings were downloaded from the Xeno-canto Archive (http://www.xeno-canto.org/). BIRDZ contains 2762 bird acoustic events with 339 detected "unknown" events that are either noise or other vocalizations aside from the eleven labeled North American bird species. Many spectrogram types (constant frequency, broadband with varying frequency components, broadband pulses, frequency modulated whistles, and strong harmonics) are included.

- CAT [41, 63]: a balanced dataset of 300 samples of ten classes of cat vocalizations in different stated collected from Kaggle, Youtube, and Flickr. The average duration of each sound is ~ 4s.
- ESC-50 [4]: an environmental sound classification dataset with 2000 samples evenly divided into fifty classes and five folds; every fold contains eight samples belonging to each class (see Table 1 for a list of the fifty classes divided by general category).

It should be noted that many papers report classification results on the datasets listed above that are superior to human performance [35-37, 39, 46, 64, 69].

| Animals | Natural soundscapes & water sounds | Human, non-speech sounds | Interior/domestic sounds | Exterior/urban noises |
|---|---|---|---|---|
| Dog | Rain | Crying baby | Door knock | Helicopter |
| Rooster | Sea waves | Sneezing | Mouse click | Chainsaw |
| Pig | Crackling fire | Clapping | Keyboard typing | Siren |
| Cow | Crickets | Breathing | Door, wood creaks | Car horn |
| Frog | Chirping birds | Coughing | Can opening | Engine |
| Cat | Water drops | Footsteps | Washing machine | Train |
| Hen | Wind | Laughing | Vacuum cleaner | Church bells |
| Insects (flying) | Pouring water | Brushing teeth | Clock alarm | Airplane |
| Sheep | Toilet flush | Snoring | Clock tick | Fireworks |
| Crow | Thunderstorm | Drinking, sipping | Glass breaking | Hand saw |

**Table 1.** Classes listed by category in the ESC-50 dataset.

The data augmentation techniques explored in this study are assessed on each dataset using the same testing protocols described in the original papers. The recognition rate (the average accuracy across all folds) is used as the performance indicator.

In Tables 2-5, the accuracy obtained by some of the data augmentation protocols is reported and compared with the baseline that skips the augmentation step (NoAUG). The CNNs were trained for 30 epochs with a learning rate of 0.0001, except for the last fully connected layer that has a learning rate of 0.001, and a batch size of 60. The one exception is the CNN labeled 'VGG16BatchSize,' the standard VGG16 with a fixed batch size of 30. For NoAUG, the batch size was set to 30.

Also, seven fusions are reported in Tables 2-5. We combined the results of the CNNs in an ensemble using the sum rule. The sum rule consists of averaging all the output probability vectors of the stand-alone CNNs in the ensemble to create a new probability vector that is used for classification. The rationale behind fusion is, as Hansen [44] describes, that "*the collective decision produced by the ensemble is less likely to be in error than the decision made by any of the individual networks*." The labels used in the tables and a brief description of the seven ensembles follow:

1. GoogleGoogle365: sum rule of GoogleNet and GoogleNetPlaces365 trained with each of the data augmentation protocols;
2. FusionLocal: sum rule of CNNs where each one is trained with a different data augmentation method;
3. FusionShort: sum rule of all CNNs trained with SGN, SSA, and SSpA;
4. FusionShortSuper: sum rule of all CNNs trained with SGN, SSA, SSpA, SSiA, and SuSA;
5. FusionSuper: sum rule of all CNNs trained with SGN, SSiA, SuSA, and TSM;
6. FusionSuperVGG16: sum rule of VGG16 trained with SGN, SSiA, SuSA, and TSM;
7. FusionALL: sum rule of all CNNs trained with SGN, SSA, SSpA, SSiA, SuSA, and TSM.

VGG16 can fail to converge; when this happens, VGG16 undergoes a second training. VGG16 can also produce a numeric problem by assigning the same scores to all patterns (random performance in the training set). In this case, all scores are considered zeros. Other numeric problems in the fusions by sum rule can occur. To avoid such issues, all scores that produce *not-a-number* value are treated as zero.

In Tables 2-4, DGT spectrogram is used for representing the signal as an image. Any cell with '---' means that the given CNN was not trained successfully (mainly due to memory problems with the GPUs).

| CAT | NoAUG | SGN | SSA | SSpA | SSiA | SuSA | TSM |
|---|---|---|---|---|---|---|---|
| AlexNet | 83.73 | 85.76 | 86.10 | 83.39 | 87.12 | 86.78 | 87.12 |
| GoogleNet | 82.98 | 86.10 | 87.80 | 83.39 | 86.78 | 85.08 | 87.80 |
| VGG16 | 84.07 | 87.12 | 88.47 | 85.76 | 87.80 | 87.80 | 88.47 |
| VGG19 | 83.05 | 85.42 | 87.80 | 84.75 | 86.10 | 86.10 | **89.15** |
| ResNet50 | 79.32 | 81.36 | 85.42 | 76.95 | 85.08 | 82.03 | 87.12 |
| ResNet101 | 80.34 | 84.75 | 85.42 | 75.59 | 82.03 | 73.56 | 86.78 |
| Inception | 79.66 | 82.71 | --- | 66.44 | --- | 84.07 | 86.10 |
| GoogleNetPlaces365 | 85.15 | 86.44 | 85.76 | 83.73 | 86.10 | 86.10 | 88.47 |
| VGG16BatchSize | --- | 86.10 | 88.14 | 86.78 | 89.49 | 86.10 | **89.15** |
| FusionLocal | 88.14 | 88.47 | 89.83 | 86.78 | 89.83 | 89.83 | **90.51** |
| FusionShort | 88.47 | | | | | | |
| FusionShortSuper | 89.83 | | | | | | |
| FusionSuper | 90.17 | | | | | | |
| FusionALL | 89.83 | | | | | | |
| FusionSuperVGG16 | 89.83 | | | | | | |

**Table 2.** Performance on the CAT dataset (reporting mean accuracy over the ten-fold cross-validation).

| BIRDZ | NoAUG | SGN | SSA | SSpA | SSiA | SuSA | TSM |
|---|---|---|---|---|---|---|---|
| AlexNet | 94.48 | 94.96 | 95.40 | 94.02 | 95.05 | 95.76 | 88.51 |
| GoogleNet | 92.41 | 94.66 | 94.84 | 91.48 | 93.85 | 95.85 | 82.91 |
| VGG16 | 95.30 | 95.59 | 95.60 | 94.69 | 95.44 | 96.18 | 94.63 |
| VGG19 | 95.19 | 95.77 | 87.15* | 94.50 | 95.44 | 96.04 | 94.88 |
| ResNet50 | 90.02 | 94.02 | 93.22 | 90.48 | 92.95 | 94.16 | 91.75 |
| ResNet101 | 89.64 | 94.00 | 92.76 | 88.36 | 92.84 | 94.20 | 90.62 |
| Inception | 87.23 | 93.84 | 92.48 | 83.81 | 92.30 | 94.01 | 90.52 |
| GoogleNetPlaces365 | 92.94 | 94.81 | 95.10 | 92.43 | 94.76 | 95.80 | 86.91 |
| VGG16BatchSize | --- | 95.84 | ---- | 94.91 | 95.81 | **96.31** | 94.78 |
| FusionLocal | 95.81 | 96.32 | 96.24 | 95.76 | 96.39 | **96.89** | 95.27 |
| FusionShort | 96.47 | | | | | | |
| FusionShortSuper | 96.79 | | | | | | |
| FusionSuper | **96.90** | | | | | | |
| FusionALL | 96.89 | | | | | | |
| FusionSuperVGG16 | 96.78 | | | | | | |

**Table 3.** Performance on the BIRDZ dataset (reporting mean accuracy over the ten-split training/test set). The * in the row VGG19 and column SSA indicates that a fold failed to converge, thus producing a random performance in that fold.

| ESC-50 | NoAUG | SGN | SSA | SSpA | SSiA | SuSA | TSM |
|---|---|---|---|---|---|---|---|
| AlexNet | 60.80 | 72.75 | 73.85 | 65.75 | 73.30 | 64.65 | 70.95 |
| GoogleNet | 60.00 | 72.30 | 73.70 | 67.85 | 73.20 | 71.70 | 73.55 |
| VGG16 | 71.60 | 79.40 | **80.90** | 75.95 | 79.35 | 77.85 | 79.05 |
| VGG19 | 71.30 | 78.95 | 78.80 | 74.10 | 78.00 | 76.40 | 77.45 |
| ResNet50 | 62.90 | 76.65 | 75.95 | 70.65 | 77.20 | 73.95 | 77.40 |
| ResNet101 | 59.10 | 75.25 | 75.65 | 70.05 | 77.50 | 72.30 | 74.85 |
| Inception | 51.10 | 71.60 | 74.70 | 63.45 | 75.55 | 71.10 | 70.65 |
| GoogleNetPlaces365 | 63.60 | 75.15 | 76.10 | 71.35 | 74.00 | 71.60 | 73.55 |
| VGG16BatchSize | --- | 79.40 | 80.50 | 73.45 | 79.35 | 77.85 | 80.00 |
| FusionLocal | 75.95 | 84.75 | 85.30 | 80.25 | 85.25 | 82.25 | 85.30 |
| FusionShort | 86.45 | | | | | | |
| FusionShortSuper | 87.15 | | | | | | |
| FusionSuper | **87.55** | | | | | | |
| FusionALL | 87.30 | | | | | | |
| FusionSuperVGG16 | 85.75 | | | | | | |

**Table 4.** Performance on the ESC-50 dataset (reporting mean accuracy over the five-fold cross-validation).

We also tested the three additional methods GA, MEL, and CO to represent a signal as an image, coupled with SGN only, as reported in Table 5.

|  | CAT | | | BIRD | | | ESC-50 | | |
| --- | --- | --- | --- | --- | --- | --- | --- | --- | --- |
|  | GA | MEL | CO | GA | MEL | CO | GA | MEL | CO |
| **AlexNet** | 82.03 | 83.73 | 79.32 | 91.85 | 91.43 | 87.54 | 73.95 | 73.50 | 65.50 |
| **GoogleNet** | 74.07 | 84.07 | 77.97 | 90.71 | 88.96 | 86.95 | 73.75 | 73.25 | 66.15 |
| **VGG16** | 83.39 | 86.10 | 80.00 | 92.65 | 93.17 | 88.82 | 77.60 | **79.20** | 66.75 |
| **VGG19** | 85.76 | 83.73 | 77.97 | 92.93 | 93.22 | 89.07 | 76.40 | 77.55 | 65.85 |
| **ResNet50** | 82.03 | 83.05 | 75.93 | 90.87 | 90.74 | 86.98 | 75.80 | 76.05 | 67.75 |
| **ResNet101** | 82.71 | 82.37 | 79.32 | 91.15 | 91.00 | 87.28 | 75.00 | 74.80 | 64.90 |
| **Inception** | 79.66 | 84.75 | 77.63 | 89.53 | 89.86 | 87.35 | 73.95 | 72.55 | 67.50 |
| **GoogleNetPlaces365** | 83.05 | 82.71 | 77.63 | 90.88 | 88.31 | 86.75 | 73.60 | 75.50 | 68.70 |
| **VGG16BatchSize** | 85.42 | **87.80** | 81.02 | 93.09 | **93.22** | 89.43 | 77.80 | 78.95 | 67.50 |
| **FusionLocal** | 87.46 | **88.47** | 82.37 | 93.76 | **93.97** | 90.57 | 81.90 | **83.80** | 73.25 |

**Table 5.** Performance using different methods for representing the signal as an image.

In Table 6, our best ensembles FusionGlobal and FusionGlobal-CO are compared with the state of the art. FusionGlobal is built with the CNNs belonging to Fusion Super and those reported in Table 5. FusionGlobal-CO is built similarly to FusionGlobal but without considering the CNNs trained using CO as a signal representation approach. The performance reported in [24] in Table 5 is different from that reported in the original paper since, for a fair comparison with this work, we ran the method without considering the supervised data augmentation approaches.

| Descriptor | BIRDZ | CAT | ESC-50 |
|---|---|---|---|
| [24] | 96.45 | 89.15 | 85.85 |
| FusionGlobal | 96.82 | 90.51 | 88.65 |
| FusionGlobal-CO | 97.00 | 90.51 | 88.55 |
| [64] | 96.3 | --- | --- |
| [2] | 95.1 | --- | --- |
| [46] | 93.6 | --- | --- |
| [63] | --- | 87.7 | --- |
| [41] | --- | 91.1 | --- |
| [41] - CNN | --- | 90.8 | --- |
| [65] | 96.7 | --- | --- |
| [32] | --- | --- | 94.10 |
| [66] | --- | --- | 89.50 |
| [33] | --- | --- | 87.10 |
| [31] | --- | --- | 88.50 |
| [40] | --- | --- | 84.90 |
| [28] | --- | --- | 86.50 |
| [29] | --- | --- | 83.50 |
| [67] | --- | --- | 83.50 |
| [30] | --- | --- | 81.95 |
| Human Accuracy [4] | --- | --- | 81.30 |

**Table 6.** Comparison of our best sound classification ensemble with state of the art.

The following conclusions can be drawn from the reported results:

1. There is no single data augmentation protocol that outperforms all the others across all the tests. TSM performs best on CAT and ESC-50 but works poorly on BIRDZ. Data augmentation at the spectrogram level works poorly on ESC-50 as well as on two other datasets. SGN and data augmentation at the signal level work well across all the datasets. On average, the best data augmentation approach is SSA. Although it produces a performance that is close to SSiA, the training time for SSA is shorter.

2. The best stand-alone CNNs are VGG16 and VGG19.

3. DGT works better than the other signal representations.

4. Combining different CNNs enhances performance across all the tested datasets.

5. For the ensemble FusionLocal, data augmentation is marginally beneficial on CAT and BIRDZ but produces excellent results on ESC-50. Compared to the stand-alone CNNs, data augmentation improves results on all three datasets. Of note, an ensemble of VGG16 (FusionSuperVGG16) outperforms the stand-alone VGG16.

6. The performance of the ensemble of CNNs trained with different augmentation policies (FusionALL) can be further improved by adding to the ensemble those networks trained using different signal representations (FusionGlobal). However, this performance improvement required considerable computation time, mainly during the training step.

7. The approach in [32] manages to outperform our results, but the authors pretrained their networks on AudioSet, hence the comparison with our approach is not fair.

The methods reported here were based solely on deep learning approaches. As mentioned in the introduction, several papers have proposed sound classification methods based on texture features. It is also possible to construct ensembles that combine deep learning with texture methods. To examine the potential of combining ensembles trained on texture descriptors with deep learning approaches, the following two fusion rules were examined:

a) Sum rule[1] between FusionGlobal and the ensemble of texture features proposed in [68] (extracted from DGT images) obtains an accuracy of 98.51% (higher than that obtained by FusionGlobal). In BIRDZ, the ensemble of texture features obtains an accuracy of 96.87%, which is close to that obtained by our deep learning approach. The ensemble with texture descriptors works poorly on ESC-50, producing an accuracy of only 70.6%. As a result, it is not advantageous to combine the texture approach with FusionGlobal.

b) The sum rule[1] between FusionGlobal and [46] obtains an excellent accuracy of 98.96% compared to [46], which achieves an accuracy of 93.6%. This means that the features extracted in [46] and by the deep learning approach access different information.

---

[1] Before the sum rule, the scores of the two ensembles are normalized to mean 0 and std 1

In terms of computation time, the most expensive activity is the conversion of audio signals into spectrograms since the conversion runs on a CPU and not on a GPU. In Table 7, computation time is reported for different CNNs and signal representations on a machine equipped with an i7-7700HQ 2.80 GHz processor, 16 GB RAM, and a GTX 1080 GPU. This test was run on an audio file of length 1.27 s with a sample rate of 32 kHz. It is interesting to note that FusionGlobal takes less than three seconds using a laptop. However, a speed-up is possible: audio files can be classified simultaneously with DGT since it can be parallelized.

| Signal Representation | Computation Time | CNN | Computation Time |
|---|---|---|---|
| **DGT** | 1.29 | **AlexNet** | 0.01 |
| **GA** | 0.02 | **GoogleNet** | 0.03 |
| **MEL** | 0.01 | **VGG16** | 0.01 |
| **CO** | 0.08 | **VGG19** | 0.01 |
| | | **ResNet50** | 0.02 |
| | | **ResNet101** | 0.03 |
| | | **Inception** | 0.03 |

**Table 7.** Computation time (in seconds) comparison between CNNs and signal representations.

## Conclusion

In this paper, we presented the largest study conducted so far that investigates ensembles of CNNs using different data augmentation techniques for audio classification. Several data augmentation approaches designed for audio signals were tested and compared with each other and with a baseline approach that did not include data augmentation. Data augmentation methods were applied to the raw audio signals and their visual representations using different spectrograms. CNNs were trained on different sets of data augmentation approaches and fusions combined by sum rule.

Experimental results clearly demonstrate that ensembles composed of fine-tuned CNNs with different architectures maximized performance on the tested three audio classification problems, with some of the ensembles obtaining results comparable with the state-of-the-art, including on the ESC-50 dataset. To the best of our knowledge, this is the largest, most exhaustive study of CNN ensembles applied to the task of audio classification.

This work can be expanded further by investigating which augmentation methods (Spectrogram Augmentation vs. Signal Augmentation) work best for classifying different kinds of sounds. We also plan to apply transfer learning using spectrograms instead of natural images. A systematic selection of augmentation approaches, e.g., by iteratively evaluating an increasing subset of augmentation techniques (as is typical when evaluating different features), would require an enormous amount of time and computation power. An expert-based approach that utilizes the knowledge of environmental scientists would be the best way of handling this challenge.

This study could also be expanded by including more datasets, which would provide a more comprehensive validation of the proposed fusions. Furthermore, there is a need to investigate the impact on performance when different CNN topologies and parameter settings in the re-tuning step are coupled with different types of data augmentation.

**Acknowledgment**

The authors thank NVIDIA Corporation for supporting this work by donating Titan Xp GPU and the Tampere Center for Scientific Computing for generous computational resources.